%% file: main.tex
\documentclass{article}
\usepackage{graphicx} 
\usepackage[hidelinks]{hyperref}
\usepackage[a4paper, left=3.5cm, right=3.5cm]{geometry}
\usepackage[backend=biber, style=apa, uniquelist=false]{biblatex}
\usepackage[british]{babel}
\usepackage{url}
\usepackage[hang,flushmargin]{footmisc}
\usepackage{hyperref}
\hypersetup{colorlinks = true, linkcolor = black, urlcolor = gray, citecolor = black, anchorcolor = black}
\usepackage{booktabs}
\usepackage{amsfonts}
\usepackage{nicefrac}
\usepackage{xcolor}
\usepackage{tabularx}
\usepackage{array}
\usepackage{acro}
\usepackage{csquotes}
\usepackage{amsmath}
\usepackage{titlesec}
\usepackage{caption}
\usepackage{changepage}
\usepackage{orcidlink}
\usepackage{cleveref}
\usepackage{float}

\titleformat*{\section}{\normalsize\bfseries}
\titleformat*{\subsection}{\normalsize\bfseries}

\titlespacing*{\section}
  {0pt}{15pt}{0pt}

\newcommand{\affilorcid}[2]{%
$^{#1,\orcidlink{#2}}$%
}

\input{helpers/acronyms}
\input{helpers/commands}

\title{EntityWeaver: Visual Exploration and Curation of Named-Entity Relationships in Document Collections \\ \vspace{0.5cm} {\large --- Technical Report ---}}

\author{
Uroš Šmajdek,\affilorcid{1}{0000-0002-8127-7700}
Ciril Bohak,\affilorcid{1}{0000-0002-9015-2897}
\vspace{0.5em}
\\
$^{1}$ University of Ljubljana, Faculty of Computer and Information Science\\
}

\date{}
                 
\begin{document}
\maketitle

\vspace{1.0cm}

\begin{abstract}
\noindent
We present an interactive visualization system for exploring named entities and their relationships across document collections, with a strong focus on handling uncertainty and supporting both distant and close reading. The system is built around a graph that links documents, entity mentions, and entities. Uncertainty from mention-to-entity linking is included directly in this graph, so users can see where connections are strong, weak, or ambiguous. A transfer-function control, inspired by approaches in scientific visualization, allows users to adjust how this uncertainty is displayed, making it easy to tune the visualization for different datasets and research questions. The system also provides direct access to the full source texts in a coordinated view, enabling quick context checks, resolving ambiguous cases, and correcting digitization errors. Exploration is further supported through a multi-stage filter query builder, mini-map navigation for large graphs, and export options for downstream analysis. By combining uncertainty-aware graph visualization with direct interaction in the source texts, the system provides a unified workflow that supports both large-scale pattern discovery and curation of named-entity-based document collections. The design choices were supported by the domain experts, who were also involved in the initial system evaluation.
\end{abstract}

\input{content/1-introduction}
\input{content/2-related-work}
\input{content/3-design-goals}
\input{content/4-system-design}
\input{content/5-evaluation}
\input{content/6-discussion}
\input{content/7-conclusion}

\section*{Acknowledgements}

This work was supported by the Slovenian Research and Innovation Agency research programme ``Digital Humanities: resources, tools and methods'' (2022--2027) [grant number P6-0436], the support of the DARIAH-SI research infrastructure, the Slovene Common Language Resources and Technology Infrastructure, CLARIN.SI, and by the project ``Large Language Models for Digital Humanities'' (2024--2027) [grant number GC-0002].

\printbibliography

\end{document}

%% file: helpers/acronyms.tex
\DeclareAcronym{ner}{
  short = NER,
  long  = named entity recognition
}
\DeclareAcronym{nel}{
  short = NEL,
  long  = named entity linking
}
\DeclareAcronym{dh}{
  short = DH,
  long  = Digital Humanities
}
\DeclareAcronym{nlp}{
  short = NLP,
  long  = Natural Language Processing
}
\DeclareAcronym{ocr}{
  short = OCR,
  long = Optical Character Recognition
}

%% file: helpers/commands.tex
\newcommand{\etal}{\emph{et al.}}
\newcommand{\eg}{\emph{e.g., }}

\newcommand{\colorframe}[1]{\fcolorbox{black}{myyellow}{\textcolor{black}{\textbf{\textsf{#1}}}}}

\definecolor{myyellow}{rgb}{0.996,1.0,0.290}

%% file: content/1-introduction.tex
\section{Introduction}
Extracting and analyzing named entities from text corpora is a fundamental step in many computational workflows in digital humanities research, journalism analysis, historical document analysis, and open-source intelligence. Recent advances in \ac{nlp} have made large-scale \ac{ner} and \ac{nel} increasingly accessible, enabling semi-automated construction of knowledge graphs from heterogeneous document collections. These representations can reveal high-level structural and temporal patterns that are difficult to detect through close reading alone.

However, entity extraction pipelines introduce multiple types of uncertainty. Ambiguous mentions, incomplete contextual evidence, conflicting metadata, and digitization errors (such as optical character recognition noise) often result in probabilistic rather than definitive entity assignments. Despite this, many existing visualization systems present extracted relationships as fixed or authoritative, suppressing uncertainty and sometimes reinforcing unreliable connections. This mismatch becomes problematic when visual interfaces are used not only for exploratory sensemaking but also for tasks such as validation, disambiguation, and refinement. In these situations, analysts must evaluate the reliability and provenance of relationships and justify corrections.

Visualization offers a way to bridge automated extraction with human oversight by revealing uncertainty in ways that aid interpretation and decision-making. Prior work has shown the value of graph visualization for exploring semantic networks and coordinated text views for reviewing source evidence. However, open challenges remain in visually encoding uncertainty in entity--mention relationships in perceptually meaningful and scalable ways, supporting smooth transitions between distant reading (overview and pattern discovery) and close reading (source-level inspection), and designing systems that explicitly support iterative curation rather than only exploratory analysis.

To address these challenges, we introduce EntityWeaver, an interactive visual analytics system for uncertainty-aware exploration and refinement of named-entity relationships across document collections. EntityWeaver represents documents, entity mentions, and resolved entities as a probabilistic multi-layer graph, displaying uncertainty directly in the visualization. Inspired by transfer-function techniques from scientific visualization, the system allows users to adjust how uncertainty is encoded to match their analytical intent—ranging from exploratory pattern analysis to conservative, validation-oriented workflows. Coordinated source text views enable analysts to examine surrounding context, resolve ambiguous cases, and correct extraction or digitization errors. Additional interaction techniques, including a multi-stage query builder, minimap-based navigation for large graphs, and export mechanisms, support iterative refinement and integration with downstream analysis pipelines.

This work makes the following contributions:
\begin{enumerate}
\item \textbf{A probabilistic multi-layer entity graph model} that incorporates uncertainty from NLP-based entity linking directly into the structure used for visual exploration.
\item \textbf{A tunable transfer-function mechanism for uncertainty encoding}, enabling users to interactively adjust the visual emphasis of uncertain relationships to match dataset characteristics and analytic goals.
\item \textbf{A unified workflow supporting both exploration and curation}, combining uncertainty-aware graph visualization with coordinated text-based verification and editing to bridge distant and close reading.
\end{enumerate}

%% file: content/2-related-work.tex
\section{Related Work}
\label{sec:related-work}
This section reviews prior work relevant to our approach. We first examine research on named entity visualization in the context of close and distant reading, followed by work on modeling and visualizing uncertainty in network graphs. These areas provide the conceptual foundation for EntityWeaver and clarify the gaps our system addresses.

\subsection{Named Entity Visualization}
Named entity visualization in the context of \acf{dh} brings together two complementary areas: named entity recognition and distant reading. Together, they enable both the extraction of structured information from textual corpora and its large-scale visual exploration.

\Acf{ner} is an \ac{nlp} task focused on automatically detecting and classifying named entities in text. The selection of entity classes depends on the domain and research objectives. Although the \emph{Person}, \emph{Location}, \emph{Organization}, and \emph{Miscellaneous} categories popularized by the CoNLL-2003 Shared Task~\cite{Tjong2003} remain widely used, the \ac{dh} community often expands these with additional types such as dates, times, products, geopolitical entities, and other domain-specific categories~\cite{Ehrmann2023}. Historically, \ac{ner} relied on handcrafted rules and feature-engineered systems, but the current state of the field is now dominated by neural network-based architectures. Keraghel~\etal~\cite{Keraghel2024} provide a comprehensive survey of these recent developments. When analyzing multi-document corpora, \ac{ner} is often supplemented by \ac{nel}, which assigns unique identifiers to individual entity instances. This normalization enables consistent tracking of entities across documents and supports downstream applications such as knowledge graph construction, metadata integration, and cross-document analysis. Guellil \etal~\cite{Guellil2024} summarize current state-of-the-art \ac{nel} methods, while Ehrmann \etal~\cite{Ehrmann2023} highlight the limitations these systems face with noisy or unconventional corpora common in \ac{dh}, particularly in historical contexts.

Distant reading, introduced by Franco Moretti in 2005~\cite{Moretti2013}, refers to analytical approaches that graph, map, or otherwise visualize textual corpora at scale. It contrasts with close reading, which focuses on the detailed examination of individual texts. Jänicke~\etal~\cite{Janicke2015} survey a wide range of visualization techniques supporting both close and distant reading and highlight a growing trend toward hybrid approaches that integrate the two. Their later work~\cite{Janicke2017} presents a detailed taxonomy of text analysis tasks and illustrates well-studied close-reading augmentations such as the use of color, glyphs, font size, and connective visual cues. The design space for distant reading is considerably broader, encompassing maps, timelines, network graphs, heatmaps, and other information visualization techniques. Among these, network visualization is the most common approach for representing named entities and the relationships between them. However, existing systems vary widely in the granularity and modalities of information they support.

A prominent example of corpus-level exploration is Serendip~\cite{Alexander2014}, a topic-centered visualization system that uses line graphs and topic-aggregation glyphs to support large-scale distant reading. While effective for analyzing topic evolution, Serendip focuses on abstract thematic structures rather than the characteristics of individual entities, making it unsuitable for tasks that require fine-grained entity-level insight. Closer to entity-based analysis are hybrid close and distant reading systems such as those by Barros Rodrigues~\etal~\cite{BarrosRodrigues2024} and El-Assady~\etal~\cite{ElAssady2017}. These works enable interplay between text context and network structure, but they are primarily designed for single-text exploration and focus mainly on network shape rather than incorporating additional entity attributes. Tamper~\etal~\cite{Tamper2023} present a more generalized approach by visualizing large collections of biographies using a combination of egocentric and sociocentric network views. Their system demonstrates the benefits of displaying entity properties such as gender and occupation. Kusnick~\etal~\cite{Kusnick2024} further extend this direction by supporting multiple entity types and providing a powerful query system for filtering a knowledge graph. However, their visualization remains centered on network structure and does not display individual entity attributes in the visual representation.

Insights from broader graph visualization applied research further emphasize the importance of coordinated multiple views, support for multiple visual modalities, and interactive filtering to facilitate exploratory analysis~\cite{Heer2005,Wang2019,Urbinati2022,Kwak2023}. While these principles are increasingly adopted in named entity visualization, they have yet to be fully leveraged for general entity-centric analyses based on user-defined entity attributes.

\subsection{Uncertainty in Network Graphs}
Uncertainty is a fundamental aspect of network models built from noisy or incomplete data. In text-derived networks, processes such as \ac{ocr}, \ac{ner}, and \ac{nel} introduce ambiguity, making relationships probabilistic. Visualizing this uncertainty is essential for accurate analysis and decision-making~\cite{Weiskopf2022,Maack2023}.

Research in uncertainty-aware visualization has produced a comprehensive set of frameworks and taxonomies for managing uncertainty throughout the visual analytics cycle. Weiskopf~\cite{Weiskopf2022} synthesizes conceptual foundations and design strategies for uncertainty visualization, emphasizing the importance of representing uncertainty at both the data and interaction levels. Maack~\etal~\cite{Maack2023} extend this perspective by defining Uncertainty-Aware Visual Analytics (UAVA), arguing that uncertainty should be propagated and made explicit throughout the entire sensemaking workflow, rather than treated as a post hoc annotation.

In network visualization, uncertainty appears in both node and edge attributes, as well as in the structure of the graph itself. One area of research focuses on uncertain network topology, where the presence or absence of edges is represented probabilistically. Schulz~\etal~\cite{Schulz2017} introduce probabilistic graph layouts that visually encode uncertainty through spatial variation, enabling analysts to reason about structural ambiguity without committing to a single deterministic representation. Related approaches sample multiple network realizations to convey structural variability, as demonstrated in the Network Hypothetical Outcome Plots technique by Zhang~\etal~\cite{Zhang2022}, which animates alternative plausible graph configurations rather than presenting a single fixed layout.

Other work explores uncertainty in node-level properties. Cesario~\etal~\cite{Cesario2011} investigate techniques for visualizing uncertainty in node attributes using glyphs and shading strategies, while more recent work introduces dynamic approaches such as animated “wiggle” encodings to visually communicate uncertainty magnitudes in a perceptually intuitive manner~\cite{Ehlers2025}. Uncertainty can also arise from downstream analytical steps such as dimensionality reduction or graph abstraction. Lan~\etal~\cite{Lan2022} address this challenge by capturing and visualizing uncertainty introduced during graph coarsening, highlighting the compounding effects of multiple processing stages.

Despite a growing body of research, applications of uncertainty-aware network visualization in digital humanities and document-centric analysis remain limited. Conroy~\etal~\cite{Conroy2023} note that humanities knowledge graphs are often treated as authoritative, despite being derived from interpretive or uncertain evidence. Their work calls for visualization systems that explicitly reveal ambiguity, support interpretive decision-making, and maintain traceability to textual sources.

EntityWeaver builds on these foundations by integrating uncertainty directly into the visual encoding of entity–mention relationships while supporting seamless transitions between uncertainty-aware distant reading and contextual close reading. Unlike systems that focus solely on network structure or uncertainty annotation, EntityWeaver embeds uncertainty into both the representation and interaction model, enabling analysts to explore, validate, and refine entity relationships grounded in source evidence.

%% file: content/3-design-goals.tex
\section{System Requirements} 
\label{sec:design_goals}
Visualizing named entities across document collections poses challenges that existing systems address only partially, especially regarding uncertainty, cross-document linking, and iterative refinement. Based on gaps identified in previous work, principles from visual analytics, and input from collaborating \ac{dh} experts, we established a set of requirements for an effective \ac{ner} visualization and curation system. These six requirements guided the design of EntityWeaver:

\begin{itemize}
  \setlength\itemsep{1em}
  \setlength{\labelsep}{1.2em}
    \item[\textbf{R1}] \textbf{Integrated data exploration and visualization --} Graph-based visualizations of named entities often require selecting a subset of entities and relationships from a much larger corpus, even when the appropriate scope or level of relevance is not known beforehand. When performed manually, this selection and refinement process can be time-consuming and difficult to iterate on. Effective systems therefore need to support flexible filtering and querying mechanisms while also enabling direct interaction with the graph itself. Such interactions include on-node and on-edge operations that allow users to modify the network in real time, thereby maintaining a fluid and exploratory workflow.
        
    \item[\textbf{R2}] \textbf{Cross-document entity relations --}
    Beyond the classification of individual mentions, \ac{nel} and coreference resolution aim to identify and connect multiple textual references that denote the same underlying entity. This challenge arises both within individual documents, where an entity may appear under different surface forms, abbreviations, or titles, and across documents, where mentions must be linked despite variation in spelling, context, or language. An effective system must account for these differences to produce coherent entity representations that span an entire corpus.

    \item[\textbf{R3}] \textbf{Extensible entity schema --} Entity datasets, especially in the context of \ac{dh}, often contain additional metadata. The system must therefore allow users to define, attach, and modify custom attributes, while also integrating them into the visualization and filtering/querying pipeline.
    
    \item[\textbf{R4}] \textbf{Retaining the uncertainty of entity data --} Information extraction processes involving named entities, such as \ac{ner} and \ac{nel}, rarely produce unambiguous results. This is particularly true for historical or otherwise heterogeneous corpora, where metadata is incomplete and textual variation is common~\cite{Ehrmann2023}. Rather than yielding a single definitive outcome, these processes often return multiple candidate entities or relations, each accompanied by a confidence score~\cite{Hu2024,Zhang2024}. Instead of preserving only the ``best guess'', our approach exposes these alternative candidates and enables researchers to examine the full range of plausible interpretations.

    \item[\textbf{R5}] \textbf{Coordinated close and distant reading --} Researchers studying named entities must continuously navigate between document-level evidence (close reading) and corpus-level patterns (distant reading). Graph view should support direct navigation to the underlying textual passages, while textual inspection should update and contextualize the visual analytics. The system must provide tight coupling between these modes of inquiry, enabling synchronized selection, highlighting, and interaction across graph visualizations, document viewers, and metadata panels.

    \item[\textbf{R6}] \textbf{Data curation --}
    Named-entity datasets derived from automatic extraction pipelines frequently require correction, refinement, and enrichment. Analysts must be able to fix incorrect entity links, resolve ambiguous cases, adjust metadata, and incorporate missing information that becomes apparent during close reading. Existing visualization systems often focus exclusively on exploration, offering limited support for modifying or validating the underlying data. To facilitate iterative improvement, the system must therefore provide mechanisms for editing entity attributes, reassigning or merging entities, correcting mention-to-entity links, and exporting curated results for downstream analysis.

\end{itemize}

The above requirements emphasize integrated exploration, cross-document entity modeling, flexible attribute handling, uncertainty retention, coordinated close and distant reading, and support for data curation.

%% file: content/4-system-design.tex
\section{System Design}
\label{sec:system-design}
EntityWeaver is an interactive, web-based system for named-entity curation and visualization, implemented using HTML, CSS, TypeScript, and the React framework. It is built around a set of coordinated views that integrate interactive text exploration, graph visualization, querying and filtering, and data curation. At its core, EntityWeaver maintains structured data models representing text, nodes, edges, and their associated metadata, supporting efficient storage, retrieval, and modification of graph information. These models drive a dedicated graph-visualization pipeline that computes layouts, applies visual styling, manages interaction behaviors, and renders the graph network for exploration and analysis. In the following subsections, we first present the data model used in our system, followed with the presentation of the visualization pipeline and the presentation of the system user interface.

\begin{figure*}[t]
    \centering
    \includegraphics[width=\linewidth]{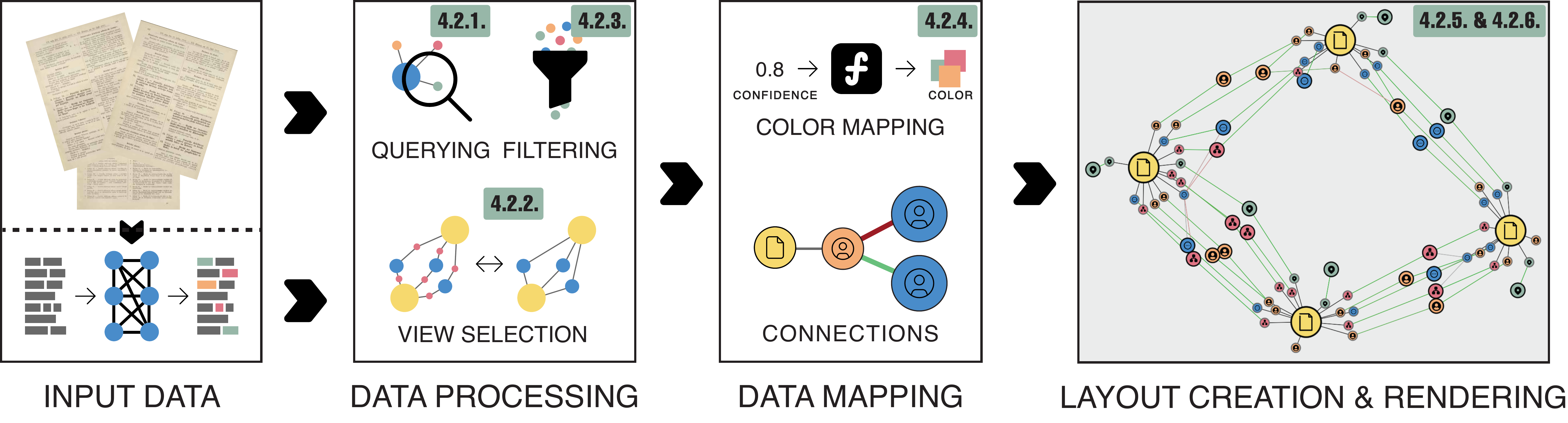}
    \caption{Pipeline for uncertainty-aware named-entity visualization in EntityWeaver. Input data is first processed through querying, filtering, and view selection. Attribute and uncertainty values are then mapped to visual encodings, after which the graph is positioned and rendered using the system’s layout and GPU-based rendering components.}
    \label{fig:vis-pipeline}
\end{figure*}

\subsection{Data Model}
\label{sec:data-model}
Graph-based named entity visualizations that depict a single document typically create entity nodes directly from entity instances within the text. Alternatively, if they are depicting some form of cross-document entity relations, \ac{nel} is first used to group those entities, in order to observe which entities appear in multiple documents. To meet the design requirements R2, R3, and R4, we chose an alternative approach and depicted both kinds of entity representations simultaneously. While related work often terms both as entities, we will define the entity instances directly obtained from a passage in a document as \emph{mentions} and the entities created through \ac{nel} as \emph{entities} in order to avoid confusion. To make cross-document relations clearer, we chose to depict the source documents as nodes as well. Our internal model thus consists of three types of nodes: \emph{mention nodes}, \emph{entity nodes}, and \emph{document nodes}.

In accordance with R3, each node can possess an arbitrary number of user-defined attributes, which may be textual, numeric, temporal, or categorical (e.g., entity type). Every attribute is associated with a unique identifier, display label, and display color; this convention also applies to categorical variables (e.g., the entity type ``Person'' has the identifier PER, display label ``Person'', and is visually represented as a blue node with an icon of a person). To support consistent graph rendering, certain attributes must always be present: document nodes require titles, mention nodes require position indices within the text, and entity nodes require names.

The data model defines three types of graph edges. The first two, \emph{mention-to-document} and \emph{mention-to-entity}, represent relationships produced by \ac{ner} and \ac{nel} and are taken directly from the input data. In line with R4, we do not restrict the number of entities linked to a single mention, allowing \ac{dh} researchers to retain all candidate associations returned by an \ac{nel} model. Each \emph{mention-to-entity} edge may include a confidence score indicating the uncertainty of the link; if no score is provided, the system assigns a default value \(c_{m-e} = 1/n\), where \(n\) is the number of entity candidates for that mention.

To support cross-document analysis and improve filtering and navigation, we additionally introduce a virtual \emph{entity-to-document} edge. This edge is derived from existing \emph{mention-to-document} and \emph{mention-to-entity} connections and links an entity to every document in which it is mentioned. Its confidence is defined as the maximum confidence among the contributing \emph{mention-to-entity} edges.

\begin{figure}[ht]
    \centering
    \includegraphics[width=0.66\linewidth]{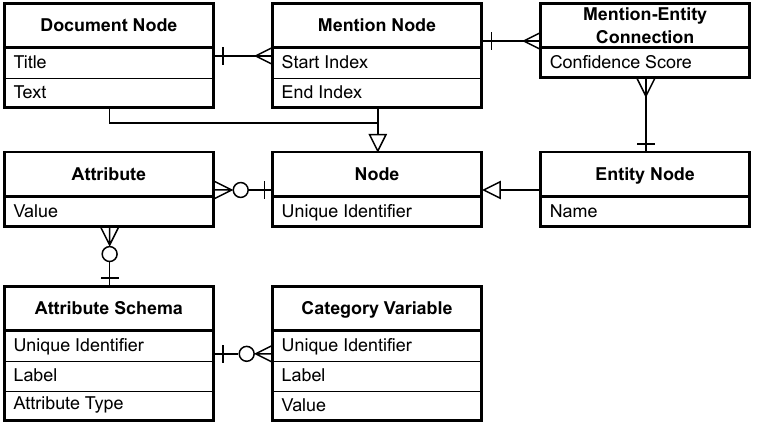}
    \caption{Entity-relation model describing EntityWeaver's internal data model.}
    \label{fig:er-diagram}
\end{figure}

An overview of the complete internal data structure is shown in~\cref{fig:er-diagram}.

\subsection{Visualization Pipeline}
\label{sec:visualization-pipeline}

The visualization pipeline, shown in \cref{fig:vis-pipeline}, is a core process in the EntityWeaver system, transforming raw data and various user inputs into a network graph. It is organized into six main stages, listed below. The pipeline is optimized to recompute a stage only when any of its preceding stages have changed, minimizing unnecessary processing.

\begin{enumerate}
    \item \textbf{Data Querying --} The input data is refined according to user-specified criteria, such as rule-based queries. node focus or node selection, to restrict the visualization to relevant subsets of the graph.
    \item \textbf{View Selection --} The structural abstraction of the graph is adjusted to present different levels of detail, enabling users to switch between high-level overviews and fine-grained representations.
    \item \textbf{Data Filtering --} Input data is refined according to user-specified criteria, such as rule-based queries, graph node focus and selection, to focus the visualization on relevant nodes and edges.
    \item \textbf{Data Mapping --} The filtered data is mapped to geometric primitives, with selected elements receiving additional visual emphasis to reflect user interactions or analytical focus.
    \item \textbf{Layout Creation --} Node and edge positions are computed to produce a spatial arrangement that highlights relational structure while preserving readability and reducing visual clutter.
    \item \textbf{Rendering --} The final graph representation is drawn on the screen, incorporating layout, styling, and interaction states.
\end{enumerate}

\subsubsection{Data Querying}
\label{sec:data-querying}
\begin{figure}[ht]
    \centering
    \includegraphics[width=0.66\linewidth]{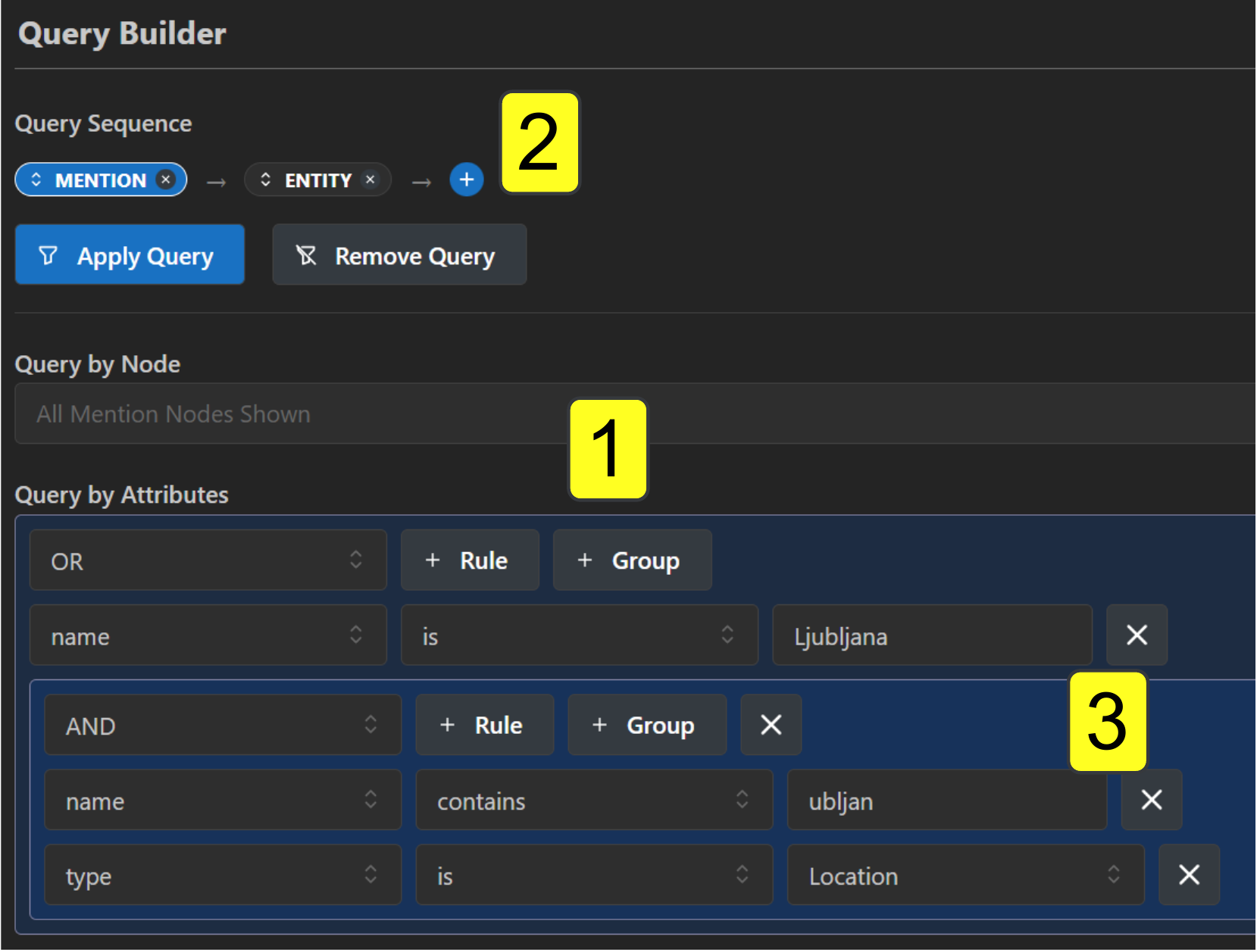}
    \caption{Data querying interface. \colorframe{\textup{1}} Query group, a set of queries applied to a single node type. \colorframe{\textup{2}} Query sequence, a list of query groups tied to different node types that are applied in the defined order. \colorframe{\textup{3}} Individual node attribute queries.}
    \label{fig:queries}
\end{figure}

During dataset analysis, users often need to narrow their focus to a specific subset of the data (R1). To support this, the system provides a rule-based querying tool. Queries are organized into \emph{query groups} (\cref{fig:queries}, \colorframe{1}), each targeting a specific node type to enable intuitive filtering across multiple types (\eg documents and mentions). These groups are arranged sequentially into a \emph{query sequence} (\cref{fig:queries}, \colorframe{2}). At execution time, the groups are evaluated in order, and the resulting set of nodes consists of those that either satisfy all query groups relevant to their node type or are connected to nodes returned by previous group in the sequence.

Each query group draws inspiration from SQL \texttt{WHERE} clauses and supports partial string matching, numerical comparisons, range queries, and categorical selections, depending on the attribute type defined by the user (\cref{fig:queries}, \colorframe{3}). Within a group, conditions can be combined using AND/OR logic blocks, allowing users to construct complex and expressive filters.

\subsubsection{View Selection}
\label{sec:view-selection}
\begin{figure}[ht]
    \centering
    \fbox{\includegraphics[width=0.35\linewidth]{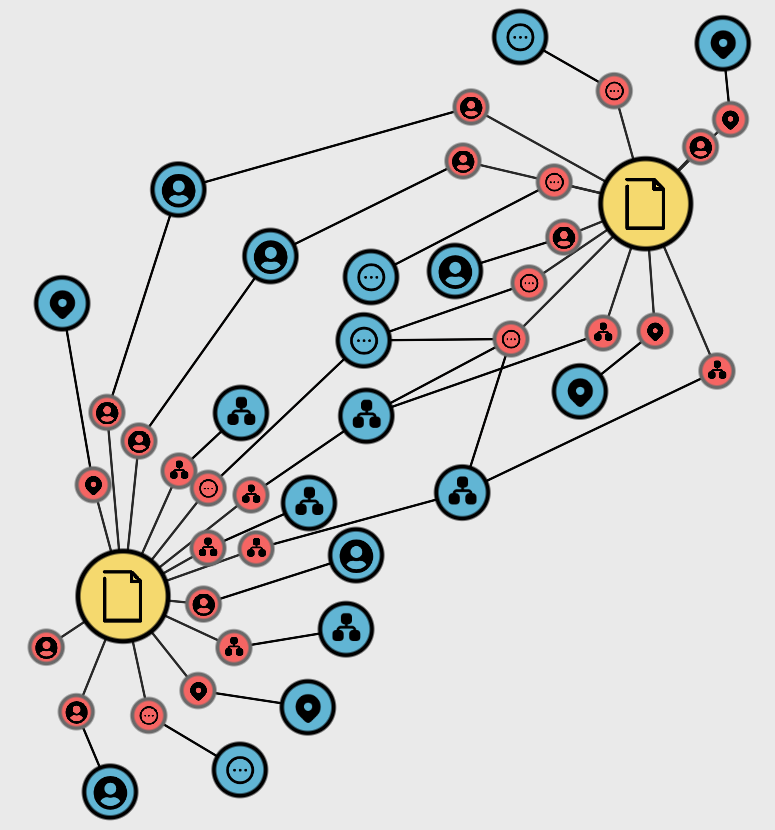}}%
    \hspace{1mm}
    \fbox{\includegraphics[width=0.35\linewidth]{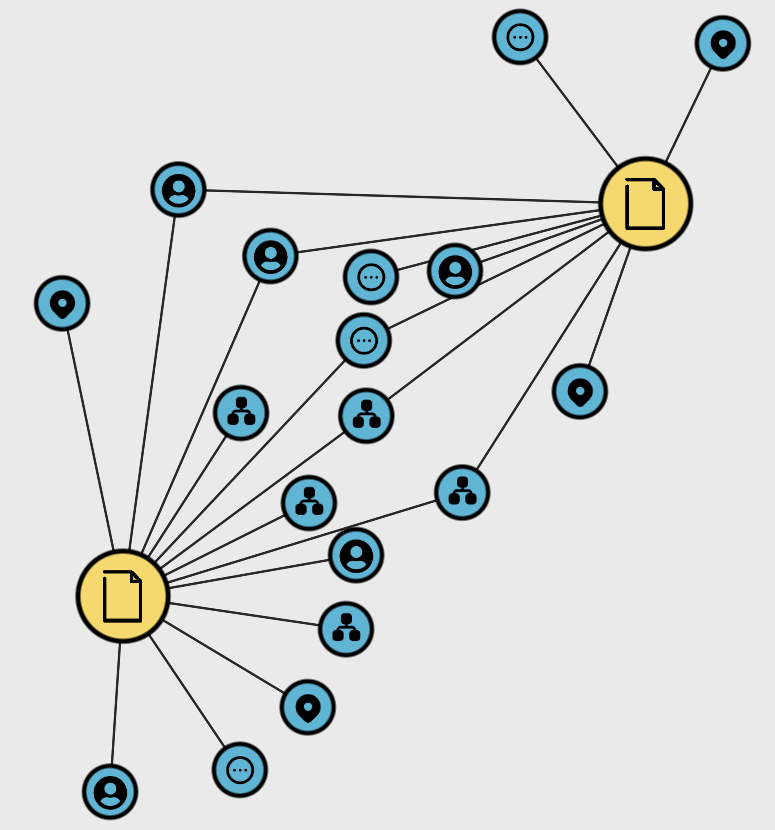}}%
    \caption{A comparison of Document–Mention–Entity view (left) and Document–Entity view (right).}
    \label{fig:DME-vs-DE}
\end{figure}

To create a more compact visualization that highlights cross-document entity relations (R4), users can switch between two structural representations of the graph: the default Document–Mention–Entity (D–M–E) view and a simplified Document–Entity (D–E) view. In the D–E view, \emph{mention nodes} are omitted and replaced by virtual \emph{entity-to-document} edges (see~\cref{sec:data-model} and \cref{fig:DME-vs-DE}).

This choice determines which nodes and edges are passed to later stages of the pipeline. The D–M–E view preserves full mention-level detail for fine-grained inspection of textual annotations, while the D–E view abstracts away mention nodes to reveal higher-level relationships across documents with reduced visual clutter.

\subsubsection{Data Filtering}
\label{sec:data-filtering}
\begin{figure}
    \centering
    \includegraphics[width=0.85\linewidth]{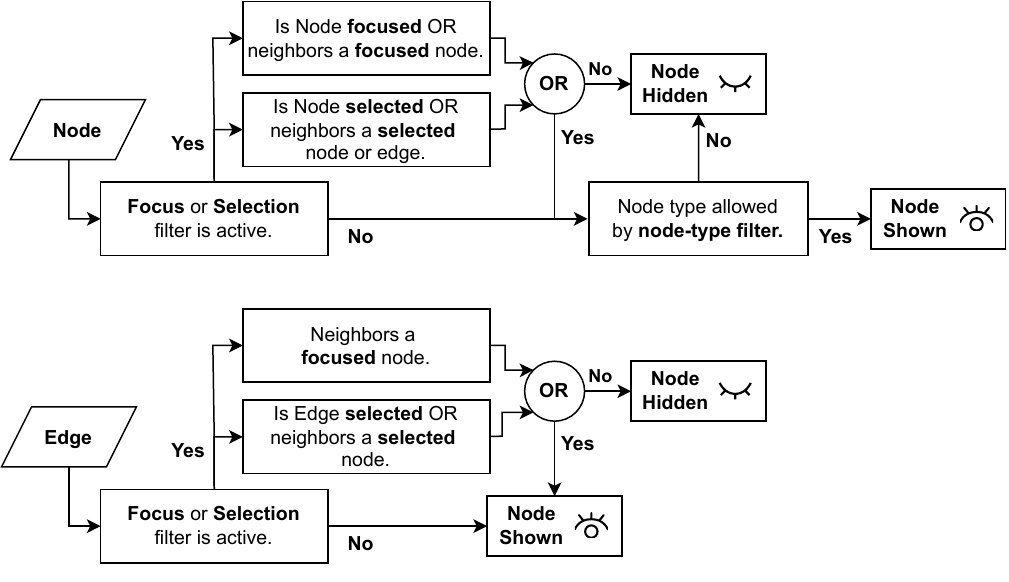}
    \caption{Node (top) and edge (bottom) filtering algorithm flowchart.}
    \label{fig:algorithms}
\end{figure}

The main goal of data filtering is to facilitate the interactivity of the graph visualization in the form of real-time responses to user input (R1). The filtering stage is composed of two sequential steps:
\begin{enumerate}
  \setlength\itemsep{1em}
    \item \textbf{Focus filter --} This filter allows the user to specify a node or edge of interest. Two focus filters can be applied concurrently: the \emph{selection filter} and the \emph{node focus filter}. The \emph{selection filter} highlights a chosen node or edge by removing all edges that do not connect it to its neighbors, allowing the user to concentrate on a specific node regardless of graph density. The \emph{node focus filter} removes all nodes and edges not directly connected to a specified node, enabling users to concentrate on the local neighborhood of interest. If applied concurrently, they are designed not to conflict with each other; the selection filter will not remove edges of the focused node and \textit{vice versa}.
    
    \item \textbf{Node-type filter --} This filter allows users to control which types of nodes (e.g., \emph{document}, \emph{mention}, \emph{entity}) are shown on the final graph.
\end{enumerate}
The complete filtering algorithm is illustrated in~\cref{fig:algorithms}.

\begin{figure}
    \centering
    \fbox{\includegraphics[width=0.30\linewidth]{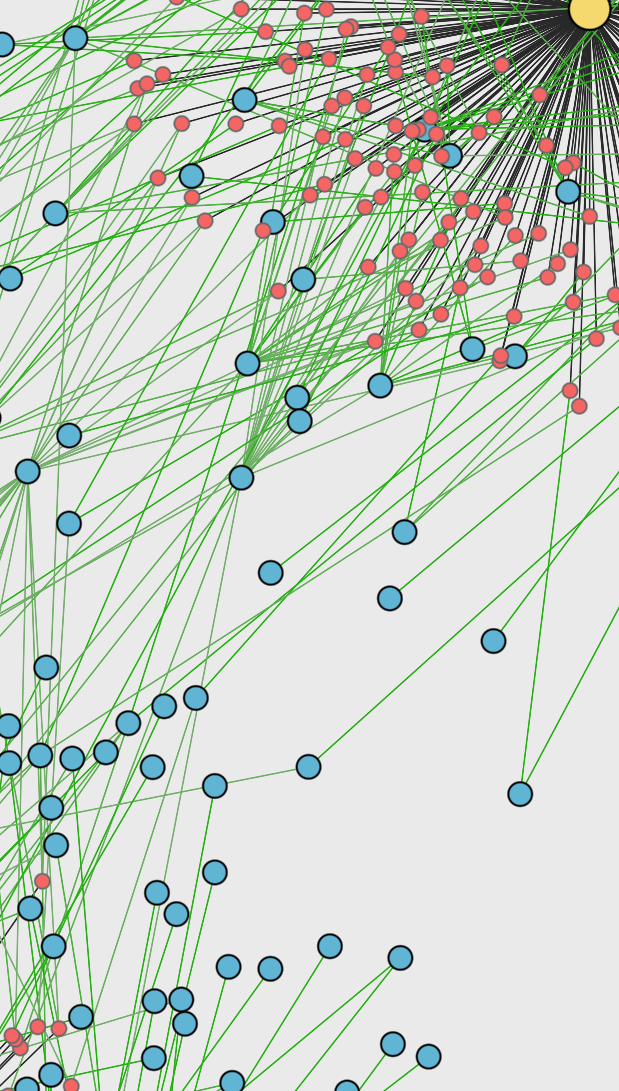}}%
    \hfill
    \fbox{\includegraphics[width=0.30\linewidth]{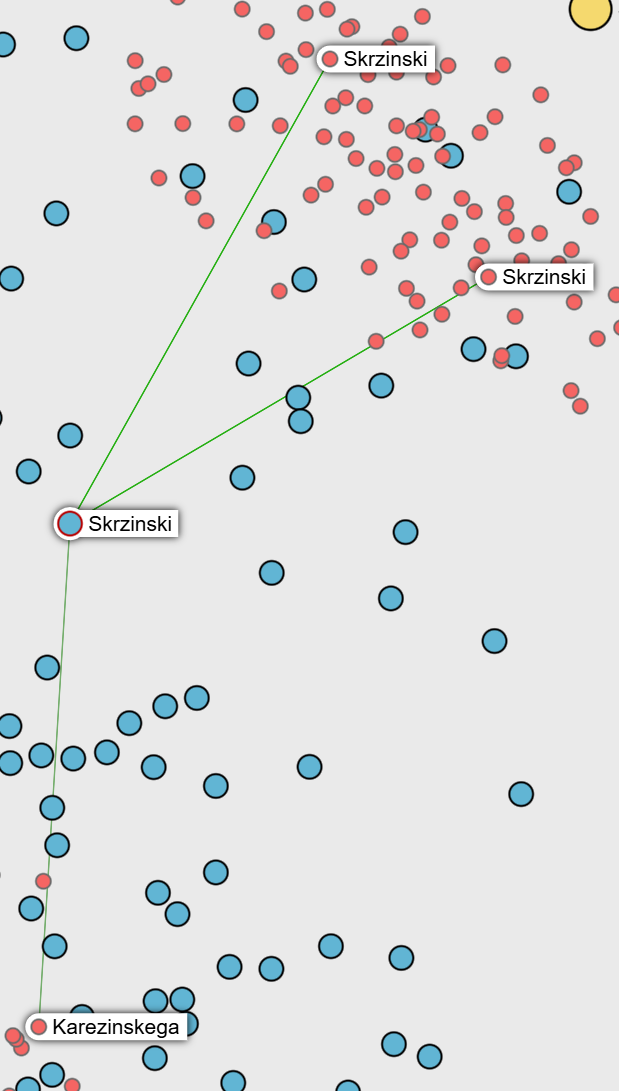}}%
    \hfill
    \fbox{\includegraphics[width=0.30\linewidth]{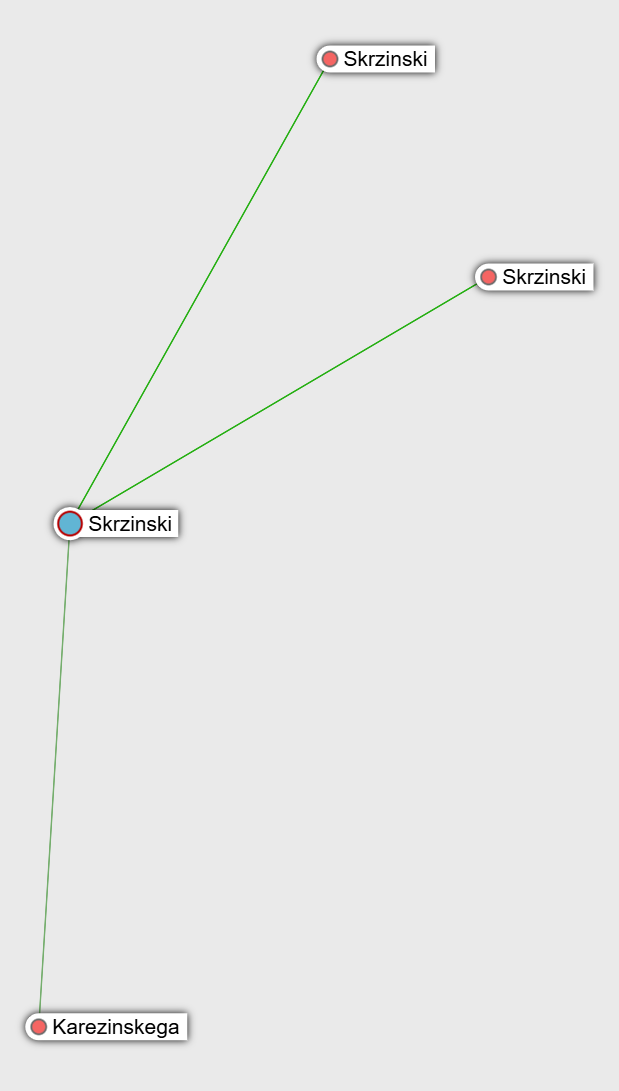}}%
    \caption{Showcase of interactive data-filtering techniques. The first panel displays an unfiltered network; the second applies a node-selection filter; and the third isolates a chosen node and its neighbours using a focus filter.}
    \label{fig:filtering}
\end{figure}

\subsubsection{Data Mapping}
\label{sec:data-mapping}
In the data mapping stage, the filtered data is transformed into circular node structures for rendering. To support the visualization of diverse node attributes (R3), we employ a three-layer node design.

\begin{figure}[ht]
    \centering
    \includegraphics[width=0.49\linewidth]{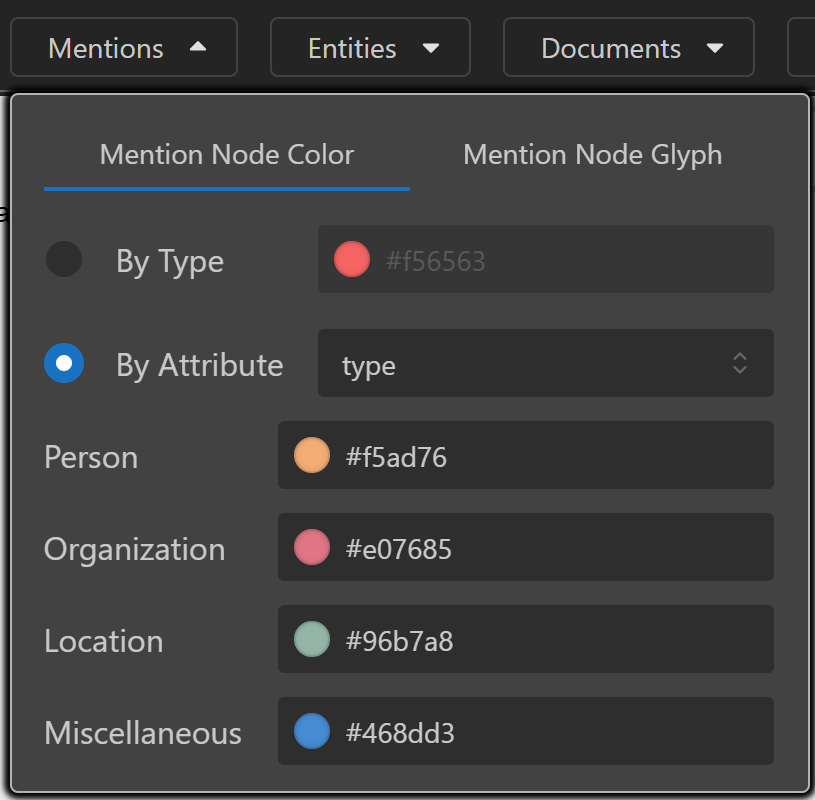}%
    \hfill
    \includegraphics[width=0.49\linewidth]{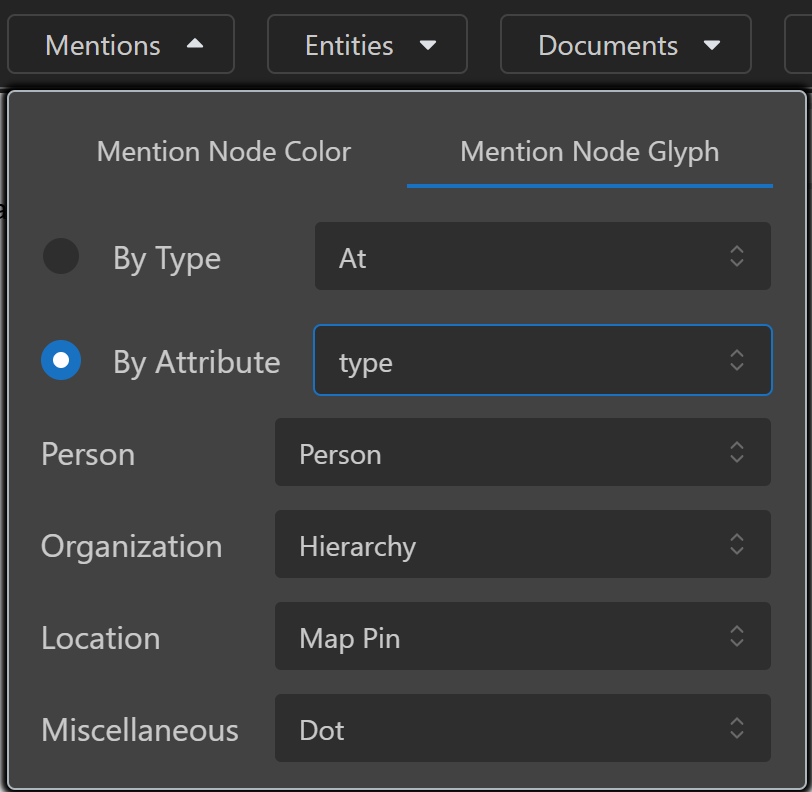}%
    \caption{Visual encoding editor, which allows manual color (left) and glyph (right) assignment. }
    \label{fig:node-editor}
\end{figure}

The first layer encodes node color, which can represent either categorical or numerical attributes. For categorical values, users may assign colors manually, with an initial palette of complementary colors provided by the system (see~\cref{fig:node-editor}, left). For numerical attributes, color is controlled through a transfer-function widget inspired by techniques from scientific visualization. This approach is particularly important in the context of named-entity analysis: it enables users to construct continuous, semantically meaningful encodings for arbitrary numerical attributes without imposing predefined interpretations. They can also be used in cases such as suppressing edges below a given confidence threshold or highlight ambiguous links. To our knowledge, such a flexible, user-driven transfer-function mechanism has not been applied in named-entity visualization before, and it significantly expands the range of analytical questions the system can support.

The second layer represents a node glyph, defined from categorical attributes (see~\cref{fig:node-editor}, right). This allows users to depict a core attribute of a node type—such as entity type—while reserving color for another value relevant to the current task.

The third layer represents the color of the node border, which is determined by the currently set focus filters (see~\cref{sec:data-filtering}). This coloring is coordinated by the depiction of the selected and focused node in other parts of the system, providing coordination between different views (R5).

To visualize the uncertainty arising from the \ac{nel} process (R4), we encoded this information into \emph{mention–entity} edges. While a variety of well-established techniques exist for depicting edge uncertainty in network graphs~\cite{Zhang2022}, we opted to use color exclusively. This decision was motivated by the potential size and density of multi-document named-entity graphs, in which modalities such as edge shape or halo become difficult to perceive. We also experimented with edge width, but found that it often obscures other graph elements and increases visual clutter due to overlap. It also allows it to be controlled using a transfer function, consistent with the depiction of numerical attributes. This, in turn, provides a clear and continuous visual encoding of edge-level uncertainty while allowing users to define any number of thresholds relevant to their analysis. The transfer function interface is shown in \cref{fig:gui}, \colorframe{6}.

\subsubsection{Layout Creation}
\label{sec:layout-creation}
In the layout creation step, node may be positioned manually or using an automated layout algorithm. For automated layouts, we use the \emph{ForceAtlas2} algorithm~\cite{Jacomy2014}, implemented through the Graphology framework~\cite{Guillaume2025}. This implementation employs the Web Workers API to parallelize computation, ensuring that the user interface remains responsive during layout processing. An example ForceAtlas2 layout is shown in~\cref{fig:gui}, \colorframe{1}.

\subsubsection{Rendering}
\label{sec:rendering}
Rendering constitutes the final stage of the visualization pipeline, responsible for producing the graph within the user interface. To ensure a smooth and responsive experience when visualizing high-density graphs, EntityWeaver integrates the \emph{Sigma.js} -- a graph rendering framework~\cite{Jacomy2025}. Sigma.js employs an instance-based WebGL rendering pipeline, enabling offloading the computationally intensive tasks to the GPU and fully leveraging modern graphics hardware for high-performance visualization.

One limitation of this rasterized rendering approach is that it does not produce vector-based outputs, which restricts the possibility of external post-processing or vector editing of the visualization. However, this is compensated by a comprehensive set of in-system editing tools, including node manipulation, advanced querying and filtering, and data curation, allowing users to adjust the visualization directly without relying on external software.

\subsection{User Interface}
\begin{figure*}
    \centering
    \includegraphics[width=\linewidth]{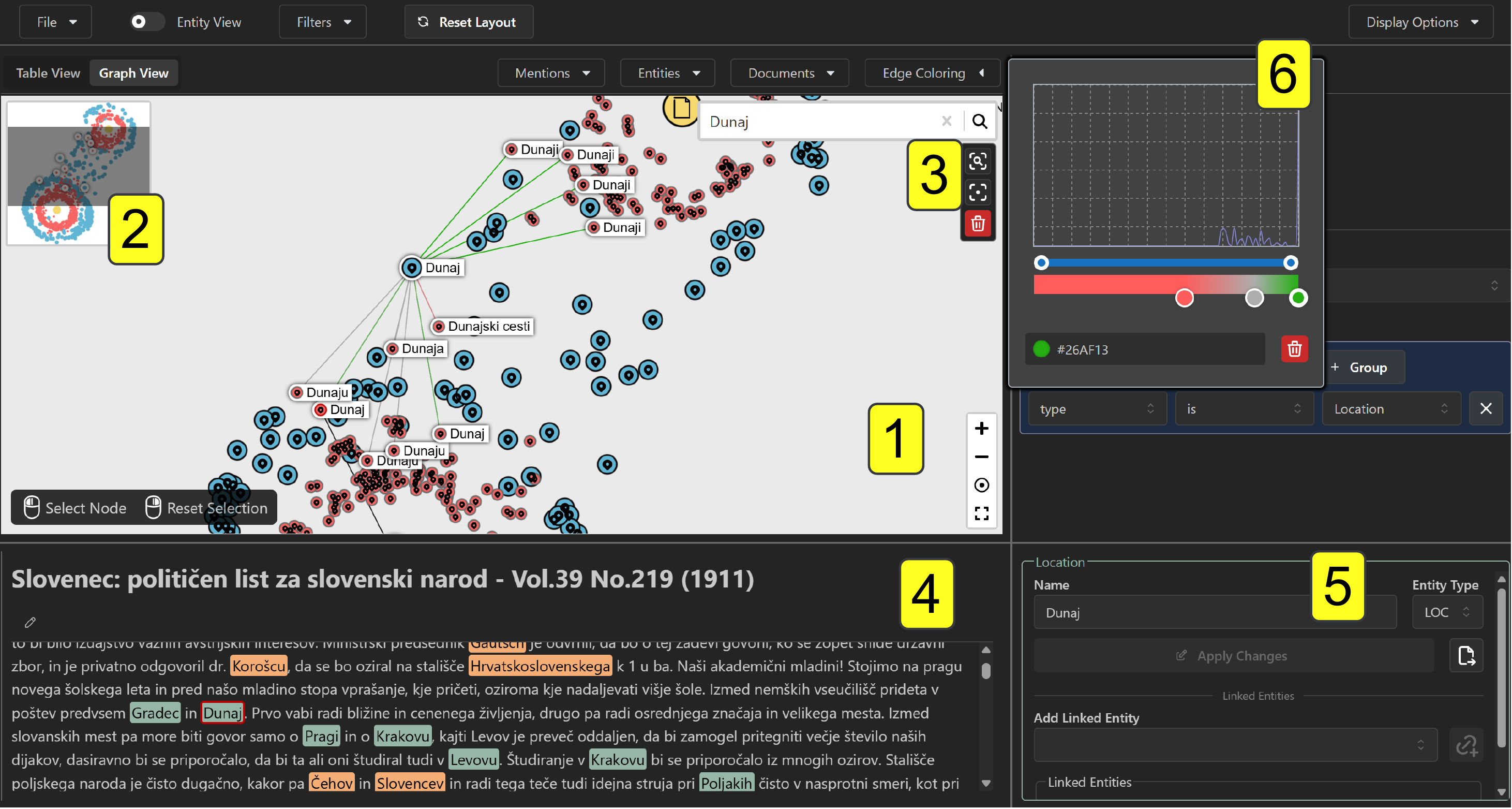}
    \caption{Overview of EntityWeaver graphical interface illustrating coordinated uncertainty-aware graph exploration and text-based curation. \colorframe{\textup{1}} Main graph view generated through the visualization pipeline (see~\cref{sec:visualization-pipeline}). \colorframe{\textup{2}} Minimap for global navigation. \colorframe{\textup{3}} Graph search with node-level actions and filters. \colorframe{\textup{4}} Source text view for close reading of selected mentions. \colorframe{\textup{5}} Node curation panel. \colorframe{\textup{6}} Transfer-function–based edge color mapper for visualizing uncertainty.}
    \label{fig:gui}
\end{figure*}

EntityWeaver is built around three coordinated views that together support both distant and close reading (R5), shown in \cref{fig:gui}. The central view presents the network graph (\cref{fig:gui}, \colorframe{1}), discussed in detail in \cref{sec:visualization-pipeline}. This view enables both a high-level overview of the network structure and a detailed inspection of individual nodes and edges. Users can directly perform operations such as selecting mentions, entities, and documents; manipulating node positions; enabling focus filters; searching through and deleting nodes (\cref{fig:gui}, \colorframe{3}). To aid navigation in larger and denser graphs, the interface also includes a minimap (\cref{fig:gui}, \colorframe{2}) that provides a compact structural overview and supports rapid repositioning within the network.

The document editor (\cref{fig:gui}, \colorframe{4}) supports close reading of selected documents (R5) as well as the textual curation and refinement of their contents (R6). The mentions are denoted directly in the text, according to their node color, as defined in the data mapping stage (see~\cref{sec:data-mapping}). The editor also integrates a two-way coordination with the network graph through synchronized node selection and hover operation: when a node is selected or hovered over in the graph, the corresponding document or mention is automatically highlighted in the editor, and \textit{vice versa}. This coordination allows the user to continuously relate textual content to the graph structure, which is especially useful when interpreting \emph{mention-entity} connections, which often require additional context to evaluate.

The node curation interface (\cref{fig:gui}, \colorframe{5}) presents the full set of metadata associated with a selected node, including all attached attributes and its incoming and outgoing connections. It also supports coordinated editing of these attributes. For attributes with visual or textual content, any modifications are immediately synchronized across the interface, ensuring that updates are reflected simultaneously in both the document editor and the network graph.

The edge color mapper (\cref{fig:gui}, \colorframe{6}), together with the other data-mapping interfaces available in the toolbar above the network graph, enables interactive adjustment of both the visualized graph elements and the annotations shown in the document editor, as described in \cref{sec:data-mapping}. To aid in transfer function design, the color mapper includes a graph showing the distribution of uncertainty values across the edges, allowing for more informed decisions. The network's toolbar also provides access to a tabular view of the dataset, offering a structured overview of all records and supporting attribute-based sorting for more systematic analysis.

%% file: content/5-evaluation.tex
\section{Expert Evaluation}
\label{sec:expert-evaluation}
As EntityWeaver is a practically oriented system, its design and development were carried out in close collaboration with a \acl{dh} scholar specializing in historical print media. This expert investigates the relationships and representations of historical figures and geographic locations within newspaper corpora that had already been processed using a variety of \ac{nlp} techniques, including \ac{ner} and \ac{nel}. He is a Ph.D.\ student in Computer Science, working at the Institute for Contemporary History, collaborating with historians and linguistics researchers. He has more than 15 years of experience with the analysis of diverse text and document corpora.

In the initial consultation, we discussed the expert’s experience with existing tools and identified key pain points. One critical insight was the high labor cost of curating high-recall, automatically generated entity sets. Historians often prefer broad candidate sets they can browse and refine manually, rather than relying on a single ``most likely'' result. This insight directly informed our implementation of R2, R4, and R6. To manage the inevitable noise from high-recall extraction, we incorporated flexible filtering mechanisms (R1). Recognizing that \ac{dh} scholars often lack engineering experience, we chose a web-based platform to simplify access and deployment.

The first round of prototype testing focused on core functionalities for data import, curation, and graph-based exploration. While feedback was genera                                       lly positive, it highlighted usability challenges in R1: deleting and modifying nodes and edges was cumbersome, and the dense, multi-document graphs made it difficult to isolate relevant information without dedicated filtering tools. These insights led to the expansion of R1 and the development of the content filtering functionality detailed in~\cref{sec:data-filtering}.

A second, more extensive evaluation took place before the presentation of the improved prototype at the HCI-SI 2025 conference~\cite{Smajdek2025}. This session highlighted persistent challenges in handling uncertainty within mention-to-entity links, prompting a refinement of R4 to emphasize clear and customizable visualization of confidence scores. The scholar also frequently needed to consult the source text to interpret ambiguous links, which motivated the formulation of R5 to support coordinated close and distant reading. Finally, the scholar requested support for additional entity types and customizable metadata fields, reinforcing the need for R3 and ensuring that the system could accommodate diverse research use cases and evolving data curation requirements.

%% file: content/6-discussion.tex
\section{Discussion}
\label{sec:fiscussion}
The feedback from the expert hints that uncertainty-aware visualization can meaningfully enhance the exploration and curation of named entities in multi-document corpora. By integrating document-, mention-, and entity-level perspectives into a unified workflow, the system supports both exploratory analysis and fine-grained validation tasks in ways not available in existing tools. The expert evaluation confirmed the importance of exposing multiple candidate entity links and providing continuous control over uncertainty encoding, reflecting the realities of historical and heterogeneous corpora where deterministic assignments are rarely sufficient.

While the collaboration with a single domain expert was instrumental in shaping the system requirements and interface design, broader evaluation is needed to fully assess the generality of the approach. Researchers in fields such as journalism, cultural analytics, or sociolinguistics may prioritize different types of entities, uncertainty characteristics, or curation workflows. Additional expert sessions, controlled task-based studies, or longitudinal deployments would provide deeper insight into system usability and applicability across domains.

Another important limitation concerns the single-user, client-side nature of the current implementation. Much humanities research is inherently collaborative, involving iterative work and shared interpretation. Extending EntityWeaver with a server-backed architecture would enable multi-user workflows, shared datasets, versioning, and conflict-resolution mechanisms. Such infrastructure would also ease the handling of large corpora that exceed browser memory limits and allow computationally intensive tasks, such as large-scale layout computation, to be offloaded to the server.

The system would also benefit from richer export capabilities. Although EntityWeaver supports basic export mechanisms, interoperability with established \ac{dh} formats such as TEI/XML, RDF/OWL, or popular knowledge-graph frameworks would allow curated results to be integrated into broader research pipelines. Exporting visualization states, including filter configurations, uncertainty thresholds, and transfer-function settings, would improve reproducibility and facilitate communication among researchers. While rasterized WebGL rendering limits the generation of vector graphics, structured data exports can serve as an effective alternative for external visualization refinement.

Scalability remains another consideration. Filtering tools and the multiple view modes help reduce visual clutter, but very large corpora can still challenge performance, particularly as users interactively refine queries or update uncertainty encodings. Server-side preprocessing, hierarchical graph abstractions, or relevance-driven sampling present promising directions for extending the system’s scalability while preserving interpretability.

Finally, EntityWeaver aligns well with emerging digital humanities workflows that emphasize the integration of distant and close reading. Embedding the system more fully into end-to-end research pipelines, such as OCR error correction, annotation environments, or web-based text repositories, would enhance its role for entity-centric analysis rather than an isolated visualization component.

%% file: content/7-conclusion.tex
\section{Conclusion}
\label{sec:conclusion}
EntityWeaver introduces an uncertainty-aware approach to exploring and curating named-entity relationships across document collections. By integrating document-, mention-, and entity-level representations, flexible filtering and querying, and coordinated close and distant reading, the system supports both large-scale analytical tasks and fine-grained interpretive work. Our collaboration with a domain expert highlighted the value of retaining alternative entity candidates, visualizing uncertainty through transfer functions, and providing editable graph structures for iterative refinement.

While the current prototype demonstrates the viability of this approach, further evaluation with a broader range of scholars, support for collaborative workflows, additional export formats, and enhanced scalability remain promising directions for future work. EntityWeaver provides a foundation for more transparent, flexible, and interpretable entity-centric analysis in digital humanities and related domains.